\documentclass[12pt]{article}

\pdfoutput=1

\usepackage[usenames,dvips]{color}
\usepackage{amssymb}
\usepackage{amsmath}
\usepackage{epsfig}
\usepackage[utf8]{inputenc}
\usepackage[T1]{fontenc}
\usepackage{graphicx}
\usepackage{xcolor}
\usepackage{tikz}

\def\d{{\rm d}}

\author{Marek Olechowski\\[12pt]
\em
Institute of Theoretical Physics, Faculty of Physics, University of Warsaw\\[6pt]
\em Pasteura 5, 02-093 Warsaw, Poland
}
\date{}
\title{\bf Stability of multibrane models}

\begin{document}
\thispagestyle{empty}

\maketitle
\begin{abstract}

Multibrane 5-dimensional models with a warped extra spatial dimension compactified on an $S^1/\mathbb{Z}_2$ orbifold are investigated. The Goldberger-Wise field, with appropriate bulk and brane potentials, is utilized to fix the brane positions. However, not all configurations with fixed brane positions are stable. 
Stability against small scalar perturbations of the metric and the Goldberger-Wise field was so far analyzed only for 2-brane models. Generalization of such analysis for multibrane models is the main subject of the present work.
Stability of brane configurations is closely related to the spectrum of radions: namely, the existence of any tachyonic radion indicates instability, while models with positive squared masses for all radions are stable. The necessary and sufficient conditions for each of these possibilities are derived for multibrane models and constitute the main result of this paper. Additionally, some properties of radions, particularly their relation to distances between and along branes, are discussed.

\end{abstract}

\maketitle

%%%%%%%%%%%%%%%%%%%%%%%%%%%%%%%%%%%%%%%%%%%%%%%%%%%%%%%%%
\section{Introduction}
\label{sec:intro}
%%%%%%%%%%%%%%%%%%%%%%%%%%%%%%%%%%%%%%%%%%%%%%%%%%%%%%%%%

Models with warped extra spatial dimensions have been intensively investigated. The famous 5-dimensional (5D) model with the 5th dimension compactified on the $S^1/\mathbb{Z}_2$ orbifold was proposed by Randall and Sundrum \cite{Randall:1999ee} as a possible solution to the hierarchy problem. In this RS model there are two 4D branes located at two fixed points of the $\mathbb{Z}_2$ symmetry. The hierarchy between Planck and electromagnetic scales exponentially depends on the distance between the two branes. 
However, in the original RS model this distance is not fixed because appropriate equations of motion (EoMs) may be satisfied for arbitrary positions of the branes. A solution of this problem was proposed by Goldberger and Wise \cite{Goldberger:1999uk} and is based on adding to the model a 5D scalar (GW scalar) with potentials in the bulk and located at the branes. This was just the beginning of very intensive investigations of models with warped orbifolds and lower dimensional branes. 
As it is stressed e.g.~in \cite{Agrawal:2022rqd}, 
they provide a frame to address fundamental questions in quantum gravity and string theory. They are also very interesting at the phenomenological level for several well known reasons (see e.g.~\cite{Davoudiasl:2009cd,Gherghetta:2000kr,Contino:2006nn,Morrissey:2009tf} and references therein).

In most models with warped extra dimensions fixing the brane positions, so also the distances between branes, is necessary to get definite predictions. However, fixing those positions is not necessarily enough to get a well defined model. It is obvious that a given configuration of branes must be stable against small perturbations. Conditions for such stability in the RS  2-brane model with GW scalar were found in \cite{Lesgourgues:2003mi}. Later analogous conditions were obtained for some more complicated models with two branes (see e.g.~\cite{Frolov:2003yi,Contaldi:2004hr,Konikowska:2006uw,Olechowski:2008bh}).

In recent years mulitbrane 5D models, i.e.~models in which there are some intermediate branes in addition to two orbifold branes, have attracted a lot of attention. More branes allow for accommodation of more hierarchies of scales. There are also more possibilities for localization of different fields on different branes. Patterns of such localization may have crucial impact on phenomenological predictions of a given model. Interactions between two fields located at different branes may be mediated only by fields propagating in the bulk (gravitons always propagate in the bulk). 
Interactions between a bulk field and a brane field depend on the profile of the bulk field and on the position of that brane.

Multibrane 5D models have been investigated with relation to several interesting theoretical and phenomenological problems (see e.g.~\cite{Oda:1999be,Oda:1999di,Hatanaka:1999ac,Kogan:1999wc,Dvali:2000ha,Gregory:2000iu,Mouslopoulos:2001uc,Kogan:2001wp,Moreau:2004qe,Agashe:2016rle,Agashe:2016kfr}).
However, the stability of such models has not been systematically studied. The main purpose of the present work is to fill in that gap. Problem of (in)stability of 5D multibrane models will be thoroughly examined. 
This problem is directly related to the spectrum of 4D modes of scalar perturbations around a given configuration of branes. Namely, existence of at least one tachyonic mode signals instability of the considered configuration. The configuration is stable if masses squared of all 4D modes are positive. The necessary and sufficient conditions for each of these possibilities will be derived and discussed.
We will see that such analysis is substantially more complicated than in the considered so far 2-brane case. One of the reasons is that one more 5D scalar perturbation must be taken into account if any intermediate brane is present which among other consequences leads to discontinuities of radion fields (absent in 2-brane models). Another reason is related to the fact that 2-brane models contain only orbifold branes for which the appropriate junction conditions are much simpler than those for intermediate branes present in multibrane models. 
Nevertheless, the results obtained in the present work may be used also for 2-brane models for which they reduce to the findings presented in \cite{Lesgourgues:2003mi}.

The problem of stability of brane configurations should not be confused with
the problem of fixing positions of the 
branes\footnote{
In the literature the notion of ``stabilization of branes'' is quite often used when in fact only the fixing of brane positions is considered. This is very misleading because solutions with such ``stabilized'' branes may very well be unstable.
}. 
The former, as mentioned above, is related to the possibility of the existence of any tachyonic 4D modes of scalar perturbations of a given background. The letter is determined by the solutions of appropriate equations of motion for that background.

%%%%%%%%%%%%%%%%%%%%%%%%%%%%%%%%%%%%%%%%%%%%%%%%%%%%%%%%

The so far most advanced investigation of stabilization of brane positions in multibrane models was presented in \cite{Lee:2021wau}. However, analysis used in \cite{Lee:2021wau} has some limitations. The authors applied the superpotential method \cite{DeWolfe:1999cp} for just one simple superpotential and also used the stiff brane potential approximation \cite{Csaki:2000zn}. In addition, only leading effects of weak backreaction of the GW field on the metric were taken into account. All those simplifications allowed to obtain e.g.~approximate formulae for the masses of two Kaluza-Klein (KK) modes of scalar perturbations (called radions).
It is not obvious how the analysis of \cite{Lee:2021wau} may be generalized for more generic bulk and brane potentials or for the case of strong backreaction. 

None of the above mentioned simplifications will be used in the present work. We will find new results related to the stability of multibrane configurations which may be applied for arbitrary bulk and brane potentials, also in the case of strong backreaction. A general Sturm-Liouville system determining the mass spectrum of radions will be given.  Exact (not only approximate) conditions necessary and sufficient for stability of multibrane configurations will be obtained. Among other results, we shall show that there may be more light radions than usually considered (e.g.~more than two light radions in models with 3 branes considered in \cite{Lee:2021wau}) and we shall discuss conditions for this to happen.

%%%%%%%%%%%%%%%%%%%%%%%%%%%%%%%%%%%%%%%%%%%%%%%%%%%%%%%%

The outline of this work is as follows. The class of models to be investigated is defined in Section \ref{sec:5Dmodels} where also a procedure of finding background solutions is presented. To find such a solution it is necessary to solve bulk Einstein equations with appropriate junction conditions (JCs) at the branes in order to obtain the background profile of the GW scalar and the metric which in turn determine the positions of all branes. 
Scalar perturbations around such background solutions are analyzed in much detail in Section \ref{sec:perturbations}. The most general relevant scalar perturbations are identified and radions are defined in terms of their 4D KK modes. Equations of motion and junction conditions for radions are found. Section \ref{sec:stability} is devoted to the analysis of the radions spectrum. Necessary and sufficient conditions for the existence of massless and tachyonic radions are derived. They allow to formulate conditions for stability of background configurations in multibrane models. Some properties of radions are discussed in Section \ref{sec:properties}. For example relations between radions and distances between branes and along branes are presented. Finally, all the results are summarized in Section \ref{sec:summary}.

%%%%%%%%%%%%%%%%%%%%%%%%%%%%%%%%%%%%%%%%%%%%%%%%%%%%%%%%%
\section{5-dimensional multibrane models}
\label{sec:5Dmodels} 
%%%%%%%%%%%%%%%%%%%%%%%%%%%%%%%%%%%%%%%%%%%%%%%%%%%%%%%%%

We consider 5D models with the action given by
\begin{equation}
\label{eq:S}
S
=
\int_{{\cal{M}}^5}{\d}^5x\sqrt{-g}
%  &
  \left[\frac{1}{2\kappa^2}R-\frac12\left(\nabla\Phi\right)^2-V(\Phi)\right.
%  \nonumber\\
%    &
    -\left.\sum_{j}\delta(y-y_j)U_j(\Phi)
    \right],
\end{equation}
where $R$ is the Ricci scalar and $\Phi$ the Goldberger-Wise scalar field. 
We assume that the 5D manifold ${{\cal{M}}^5}$ is a warped product of 4D Minkowski manifold and 1D orbifold:
\begin{equation}
\label{eq:M5}
{\cal{M}}^5={\cal{M}}^4 \times S^1/\mathbb{Z}_2\,.
\end{equation}
There are several potentials in the action \eqref{eq:S}.
$V(\Phi)$ is the bulk potential while $U_j(\Phi)$ are potentials localized
at branes positioned at some values of the 5th coordinate $y=y_j$. There are two orbifold branes at the fixed points of the $\mathbb{Z}_2$ symmetry. We denote their 5th coordinates as $y_1$ and $y_2$. In addition there may be some number of intermediate branes located at $y_{I_i}$. In a model with $N$ branes the sum over $j$ in \eqref{eq:S} is over all $N$ branes: 2 orbifold branes and $N-2$ intermediate ones.  Let us number the branes such that $y_1<y_{I_1}<y_{I_2}<\ldots<y_{I_{N-2}}<y_2$. One should keep in mind that the bulk potential $V(\Phi)$ may have different form in different interbrane sections of the bulk and may be written as
\begin{equation}
\label{eq:V_bulk}
V(\Phi)=
\begin{cases}
V_{1I_1}(\Phi) \qquad\text{for}\quad y\in(y_1,y_{I_1}) \\
V_{I_1I_2}(\Phi)  \,\,\,\,\quad\text{for}\quad y\in(y_{I_1},y_{I_2}) \\
\ldots \\
V_{I_{N-1}2}(\Phi) \quad\text{for}\quad y\in(y_{I_{N-2}},y_2)
\end{cases}
\end{equation}
In fact this is the case in most (if not in all) models considered in the literature.

We do not include extrinsic curvature
terms in the action \eqref{eq:S}, because we work in the so-called upstairs picture of the $S^1/\mathbb{Z}_2$ orbifold, where all total derivatives integrate to zero. The boundary conditions (BCs) at the orbifold branes are obtained by integrating the equations of motion around those branes.

The schematic picture illustrating the main ingredients of a multibrane model is shown in Fig.~\ref{fig}. A comment about the number of branes in a given multibrane model is in order. As is usual in the literature we will use the name ``$N$-brane model'' for models with 2 orbifold branes, $N-2$ intermediate branes in the bulk and $N-2$ branes which are $\mathbb{Z}_2$ images of those intermediate branes. There are $2N-2$ branes on the full circle $S^1$ but $N-2$ of them do not correspond to any independent degrees of freedom, so the name ``$N$-brane model'' is justified\footnote{$N$ denotes the number of independent branes and not their dimensionality, all the branes are $(3+1)$-dimensional.}.

\begin{figure*}
\begin{center}
\begin{tikzpicture}
\draw[dashed, thick] (0,0) -- (5,0);
\draw[black, thick] (5,0) -- (10,0);
\draw[red, ultra thick] (5,-0.01) -- (5,3);
\draw[blue, ultra thick] (0,-0.01) -- (0,1);
\draw[blue, ultra thick] (10,-0.01) -- (10,1);
\draw[green, ultra thick] (8.5,-0.01) -- (8.5,1.6);
\draw[green, dashed, ultra thick] (1.5,-0.01) -- (1.5,1.6);
\draw[gray, thick] (5,3) .. controls (6,2.2) and (7.5,1.7) .. (8.5,1.6);\draw[gray, dashed, thick] (5,3) .. controls (4,2.2) and (2.5,1.7) .. (1.5,1.6);
\draw[gray, thick] (8.5,1.6) .. controls (9,1.2) and (9.7,1.05) .. (10,1);
\draw[gray, dashed, thick] (1.5,1.6) .. controls (1,1.2) and (0.3,1.05) .. (0,1);
\node at (5.0,-0.3) {$y_1$};
%\node[gray] at (0,-0.3) {$\sim\!y_2$};
\node at (0.1,-0.3) {$y_2$};
\node[gray] at (1.5,-0.3) {$\sim\!y_I$};
\node at (8.5,-0.3) {$y_I$};
\node at (10,-0.3) {$y_2$};
\node at (5,3.3) {$U_1$};
\node at (8.5,1.9) {$U_I$};
\node[gray] at (1.5,1.9) {$U_I$};
\node at (10,1.3) {$U_2$};
\node at (0,1.3) {$U_2$};
\node at (6.75,0.8) {$V$};
\node at (6.75,0.3) {$(V_{1I})$};
\node at (9.25,0.8) {$V$};
\node at (9.25,0.3) {$(V_{I2})$};
\node[gray] at (3.25,0.8) {$V$};
\node[gray] at (3.25,0.3) {$(V_{1I})$};
\node[gray] at (0.75,0.8) {$V$};
\node[gray] at (0.75,0.3) {$(V_{I2})$};
\node at (9.6,3.0) {$e^{-A(y)}$};
\draw[gray, thick, ->] (9,2.7) -- (6.9,2.15);
\draw[gray, thick, ->] (9.3,2.7) -- (9.3,1.3);
\draw[gray, dashed, thick] (14.5,1.4) arc (0:180:1);
\draw[thick] (14.5,1.4) arc (0:-180:1);
\node[blue] at (14.5,1.4) {$\bullet$};
\node[red] at (12.5,1.4) {$\bullet$};
\node[green] at (14.1,0.6) {$\bullet$};
\node[green] at (14.1,2.2) {$\circ$};
\node[green] at (14.1,2.2) {$\cdot$};
\node[thick] at (13.85,1.4) {$\mathbb{Z}_2$};
\draw[ <->] (14.1,0.8) -- (14.1,2.0);
\node at (12.2,1.4) {$y_1$};
\node at (14.8,1.4) {$y_2$};
\node at (14.35,0.4) {$y_I$};
\end{tikzpicture} 
\end{center}
\caption{The schematic picture illustrating the main ingredients of a 3-brane model. Two orbifold branes (red and blue) are located at two fixed points of the $\mathbb{Z}_2$ symmetry ($y_1$ and $y_2$, respectively). The intermediate brane (green) and its $\mathbb{Z}_2$ image (dashed green) are located in the bulk and in the bulk image, respectively.}
\label{fig}
\end{figure*}

%%%%%%%%%%%%%%%%%%%%%%%%%%%%%%%%%%%%%%%%%%%%%%%%%%%%%%%%%%%%%%%%%
\subsection{Background solutions}
\label{sec:background}
%%%%%%%%%%%%%%%%%%%%%%%%%%%%%%%%%%%%%%%%%%%%%%%%%%%%%%%%%%%%%%%%%

First one has to find the background configuration for a given model. We are interested in backgrounds which may be described by just two functions of the 5th coordinate: the GW scalar field
\begin{equation}
\label{eq:Phi_ansatz}
\Phi(x^\mu,y) = \phi(y)\,,
\end{equation}
and the warp factor function $A(y)$ determining the 5D metric:\footnote{
We use the conformally flat 5D coordinates, because they are more suitable for some of the calculations, but all the results may be easily translated to often used coordinates in which only 4D metric depends on the 5th coordinate:
$\d s^2 
= e^{-2\widetilde{A}(z)}\eta_{\mu\nu}\d x^\mu \d x^\nu + \d z^2$.
}
\begin{equation}
\label{eq:metric_ansatz}
\d s^2 
%= g_{MN}\d x^M \d x^N 
= e^{-2A(y)}\left(\eta_{\mu\nu}\d x^\mu \d x^\nu + \d y^2\right)\,.
\end{equation}
In order to find these two functions one has to solve appropriate bulk equations of motion and junction/boundary conditions. The bulk EoMs for one interbrane section i.e.~for a subspace of \eqref{eq:M5} with the 5th coordinate $y$ belonging to an open interval between positions of two neighboring branes, are as follows ($V$ may be a different function of $\phi$ in different sections of the bulk):
\begin{eqnarray}
\label{eq:EoM_A''}
A''+(A')^2-\frac{\kappa^2}{3}(\phi')^2 &=& 0\,,
\\
\label{eq:EoM_phi''}
\phi''-3A'\phi'-e^{-2A}\,V'&=&0\,,
\\
\label{eq:EoM_A'phi'}
(A')^2-\frac{\kappa^2}{12}(\phi')^2 + \frac{\kappa^2}{6}e^{-2A}\,V &=& 0\,,
\end{eqnarray}
where prime denotes derivative with respect to an appropriate argument, i.e.~with respect to the 5th coordinate $y$ in case of $A(y)$ and $\phi(y)$ and with respect to the scalar field $\phi$ in case of the bulk potential $V(\phi)$ and brane potentials $U_j(\phi)$ - which appear later. 
The above equations are not independent. As usually, the equation which follows from the variation of the action with respect to the scalar field (eq.~\eqref{eq:EoM_phi''} in our case) may be obtained from the Einstein equations (in our class of models different components of the tensor Einstein equation lead to \eqref{eq:EoM_A''} and \eqref{eq:EoM_A'phi'}).
Nevertheless, each of the above equations \eqref{eq:EoM_A''}-\eqref{eq:EoM_A'phi'} will be useful in some situations.

Equations \eqref{eq:EoM_A''}-\eqref{eq:EoM_A'phi'} may be applied only for $y\ne y_j$ i.e.~away from all branes. Of course, variation of the action \eqref{eq:S} gives EoMs valid in the whole space-time. It is convenient to integrate such full 5D EoMs over an infinitesimal interval of the 5th coordinate containing the position of one brane. This procedure gives junction conditions which must be satisfied at that brane. In the considered class of models such JCs for the brane located at $y=y_j$ read
\begin{eqnarray}
\label{eq:JC_A'}
\left[A'\right]_j&=&\frac{\kappa^2}{3}\,e^{-A}\,U_j\,,
\\
\left[\phi'\right]_j&=&e^{-A}\,U_j'\,,
\label{eq:JC_phi'}
\end{eqnarray}
where $[X]_j:=X(y_j^+)-X(y_j^-)$ denotes a possible jump of quantity $X$ across the $j$-th brane. The fields $A(y)$ and $\phi(y)$ are continuous but usually not differentiable at $y_j$. Jumps of their differentials are determined by the corresponding brane potential $U_j(\phi)$ and its derivative. The above JCs simplify for the orbifold branes at $y_1$ and $y_2$ because $A'$ and $\phi'$ are functions odd under the $\mathbb{Z}_2$ symmetry and for any such function $X$ we have $[X]_{i}=2X(y_{i}^+)=-2X(y_{i}^-)$ for $i=1,2$. 
The upstairs approach to orbifolds is adopted in this work so the JCs at the $\mathbb{Z}_2$ fixed points take the form of the following boundary conditions:
\begin{eqnarray}
\label{eq:BC_A'_12}
\lim_{y\to y_i^{\pm}} A' &=& \pm \frac{\kappa^2}{6}\,e^{-A}\,U_i\,,
\\
\label{eq:BC_phi'_12}
\lim_{y\to y_i^{\pm}} \phi' &=& \pm \frac12 \,e^{-A}\,U'_i\,,
\end{eqnarray}
with upper (lower) signs for $i=1$ ($i=2$).

Let us now describe in some detail a possible procedure of constructing the background solution for a given model. It is convenient to start at one of the orbifold branes, e.g.~the one located at $y_1$. It is necessary to solve two second order differential bulk EoMs \eqref{eq:EoM_A''} and \eqref{eq:EoM_phi''}. So, to start the integration one needs initial values of four quantities: $A$, $A'$, $\phi$ and $\phi'$ all evaluated at $y_1^+$. The value of $A(y_1)$ may be chosen arbitrary, e.g.~as  $A(y_1)=0$, because this only determines our units in which we will express later the results. The initial values of the remaining three quantities at $y_1^+$ can be found by solving the set of three equations: two BCs \eqref{eq:BC_A'_12} and \eqref{eq:BC_phi'_12} with $i=1$ and the bulk equation \eqref{eq:EoM_A'phi'}. This set of equations in general has a discrete set of solutions.
Each solution corresponds to one set of allowed boundary (initial) values for our two differential bulk EoMs. With well defined initial conditions we  may integrate (usually only numerically) equations \eqref{eq:EoM_A''} and \eqref{eq:EoM_phi''}. This way we obtain two functions $A(y)$ and $\phi(y)$, and of course their derivatives, for $y>y_1$.

Now one has to place the first intermediate brane at some, yet unknown, position $y_{I_1}$. It is not possible to place it at arbitrary position.
Let us see what happens if one tries to put that brane at some position $y$. Solution of the bulk differential EoMs give us values of $A$, $A'$, $\phi$ and $\phi'$ at chosen $y^-$, i.e.~``just before'' the brane. Values of these quantities ``just after'' the brane, at $y^+$, are also know. $A$ and $\phi$ are continuous while values of $A'$ and $\phi'$ are determined by JCs \eqref{eq:JC_A'} and \eqref{eq:JC_phi'}. However, all these quantities at $y^+$ must fulfill equation \eqref{eq:EoM_A'phi'} with the bulk potential $V$ present in the second interbrane section of the space-time. Using equations \eqref{eq:EoM_A'phi'}-\eqref{eq:JC_phi'} one gets the following condition
\begin{equation}
e^{A(y)}\left(4A'(y^-)U_{I_1}(\phi(y))-\phi'(y^-)U_{I_1}^\prime(\phi(y))\right)
+
\frac{2\kappa^2}{3}\,U_{I_1}^2(\phi(y))-\frac12\left(U_{I_1}'(\phi(y))\right)^2
+[V(\phi(y))]_{I_1}=0\,.
\label{eq:brane_position}
\end{equation}
The last terms on the l.h.s.~is the difference of bulk potentials at two 
bulk sections separated by the first intermediate brane. The argument of the two potentials, $\phi(y)$, is the same but the functional form may be, and usually is, different, so in general $[V(\phi(y))]_{I_1}\ne0$.
The first intermediate brane may be placed only at such $y=y_{I_1}$ at which the equation \eqref{eq:brane_position} is fulfilled with $A$, $A'$, $\phi$ and $\phi'$ obtained from the integration of the bulk EoMs \eqref{eq:EoM_A''} and \eqref{eq:EoM_phi''} in the first interbrane section of the 
bulk.

After finding the allowed position of the first intermediate brane we repeat the procedure for the next intermediate brane. We solve the bulk EoMs in the second interbrane section and look for the position of the next brane i.e.~look for $y$ at which condition \eqref{eq:brane_position} with ${I_1}$ replaced with ${I_2}$ is fulfilled. Then we continue until the position of the last intermediate brane is 
determined\footnote{
If for any intermediate brane $I_k$ the condition \eqref{eq:brane_position} with ${I_1}$ replaced with ${I_k}$ has no solutions then the model with chosen bulk and brane potentials is inconsistent with our ansatz \eqref{eq:metric_ansatz} i.e.~it does not allow background solutions with flat Minkowski 4D sections.}. 
Finally we have to find the position of the second orbifold brane. Here the situation is different. As in the case of intermediate branes the brane position is the only parameter which may be adjusted but instead of one condition \eqref{eq:brane_position} two BCs, \eqref{eq:BC_A'_12} and \eqref{eq:BC_phi'_12} with $i=2$, must be satisfied. Thus, one fine tuning of parameters is necessary in order to get any solution. This is exactly the same situation as in 2-brane models. One tuning is necessary to have vanishing effective 4D cosmological constant in agreement with the ansatz \eqref{eq:metric_ansatz}.

Let us summarize the result of the above described procedure. 
Models are defined by specifying the number of intermediate branes and all the potentials present in the action \eqref{eq:S}: brane potentials localized at each brane and bulk potentials at each interbrane section. Background solutions of the form \eqref{eq:Phi_ansatz}-\eqref{eq:metric_ansatz} exist if all the bulk and brane equations \eqref{eq:EoM_A''}-\eqref{eq:JC_phi'} may be solved. Those equations are not all linear so there may be more than one solution for a given model. For each background solution the positions of all branes are fixed. Thus, also all interbrane distances are fixed. This is different than in the original RS model (or similar models without the Goldberger-Wise scalar field) in which brane positions are not fixed because the appropriate EoMs and BCs have solutions for arbitrary brane positions.

For any background solution of any model with branes and GW 5D scalar considered in this work the positions of all branes are fixed. However, it is not guarantied that such background in a given model is stable. In order to check whether this is the case one has to investigate small perturbations around the background. This is the subject of the next Section.

%%%%%%%%%%%%%%%%%%%%%%%%%%%%%%%%%%%%%%%%%%%%%%%%%%%%%%%%%%%%%
\section{Scalar perturbations}
\label{sec:perturbations}
%%%%%%%%%%%%%%%%%%%%%%%%%%%%%%%%%%%%%%%%%%%%%%%%%%%%%%%%%%%%%

We have shown that in models with a GW scalar field typically there are background solutions of all EoMs for which the positions of all branes are fixed\footnote{In some cases there may be a discrete set of solutions, in some other cases there may be no solutions.
}. Now we will check whether such branes are stabilized at those fixed positions. In order to do this we have to generalize the background ansatz \eqref{eq:Phi_ansatz}-\eqref{eq:metric_ansatz} by adding small scalar perturbations. The generalized ansatz reads:
\begin{align}
\label{eq:metric_ansatz_perturb}
\d s^2 
&= e^{-2A(y)}\left\{
\left[\left(1+2F_1(x,y)\right)\eta_{\mu\nu}
+\partial_\mu\partial_\nu E(x,y)\right]\d x^\mu \d x^\nu 
 + \left(1+2F_2(x,y)\right)\d y^2\right\}\,,
\\
\label{eq:Phi_ansatz_perturb}
\Phi &= \phi(y)+F_3(x,y)\,.
\end{align}
In the case of 2-brane models it is enough to consider only three 
perturbations $F_k(x,y)$, $k=1,2,3$. It is necessary to add yet another metric perturbation, $E(x,y)$, if any intermediate brane is added 
\cite{Pilo:2000et,Kogan:2001qx}. 
All these perturbations are 5D fields depending on all coordinates. 
However, they are not independent. The off-diagonal components of tensor Einstein EoM give some relations between the scalar perturbations. From the $(\mu,5)$ components it follows that 
\begin{equation}
\label{eq:F3F2F1}
3F_1'+3A'F_2+\kappa^2\phi'F_3 = 0\,,
\end{equation}
while $(0,i)$ components, after using diagonal EoMs \eqref{eq:EoM_A''} and \eqref{eq:EoM_phi''}, give the equation
\begin{equation}
\label{eq:F2F1}
2F_2 +4F_1- E^{\prime\prime}+3A'E' = 0\,.
\end{equation}
The last two equations may be used to express two perturbations in terms of the remaining two (e.g.~$F_2$ and $F_3$ in terms of $F_1$ and $E$). It is not difficult to check that the bulk EoM for the perturbation $E$ is automatically fulfilled when the background EoMs \eqref{eq:EoM_A''}-\eqref{eq:EoM_A'phi'} are used. Thus, in the bulk we are left with just one independent perturbation. Let us define it as the following combination:
\begin{equation}
\label{eq:F_def}
F:=F_1+\frac12A'E'\,.
\end{equation}
The equation of motion for $F$ follows from the (5,5) component of the Einstein equation for the metric \eqref{eq:metric_ansatz_perturb} and reads
\begin{equation}
\label{eq:EoM_F}
F^{\prime\prime}-\left(3A'+2\frac{\phi^{\prime\prime}}{\phi'}\right)F'
+\left(4A'\frac{\phi^{\prime\prime}}{\phi'}-2A^{\prime\prime}+2\left(A'\right)^2-\frac23\kappa^2\left(\phi'\right)^2-\Box_4\right)F=0\,,
\end{equation}
where $\Box_4$ is the 4D d'Alambertian.

The importance of the scalar perturbation $E$ goes beyond the contribution of $E'$ to the definition \eqref{eq:F_def} of $F$. 
Although all bulk EoMs involving $E$ other than \eqref{eq:EoM_F} are automatically fulfilled, values and jumps of derivatives of $E$ at the branes do play important role. Such jumps may be derived from equation \eqref{eq:F2F1}. Scalar perturbations of the metric \eqref{eq:metric_ansatz_perturb}, $F_1$ and $F_2$, must be continuous (also at the brane locations) while $A'$ must be finite (although it may be discontinuous). Taking this into account one finds that at all branes $E'$ must be continuous while the jump of $E^{\prime\prime}$ is proportional to that of $A'$:
\begin{align}
\label{eq:E'}
[E']_j&=0\,,
\\
\label{eq:E''}
\left[E^{\prime\prime}\right]_j&=3E'(y_j)[A']_j\,.
\end{align} 

As we have shown, only one combination of the four scalar perturbations introduced in eqs.~\eqref{eq:metric_ansatz_perturb} and \eqref{eq:Phi_ansatz_perturb} is independent. We have chosen $F$ defined in \eqref{eq:F_def} but it is convenient to modify that definition by multiplying it by the warp factor. Thus, from now on we will use the following 5D scalar field
\begin{equation}
\label{eq:Q_def}
Q(x^\mu,y):=e^{-2A(y)}\left(F_1(x^\mu,y)+\frac12A'(y)E'(x^\mu,y)\right).
\end{equation}
One should notice that $Q$ (as well as $F$ defined in \eqref{eq:F_def}) may be discontinuous at the branes. The metric perturbation $F_1$ must be continuous so the jump of $Q$ at the position of the $j$-th brane equals
\begin{equation}
\label{eq:[Q]}
\left[Q\right]_j=\frac12 e^{-2A(y_j)} E'(y_j)\left[A'\right]_j\,.
\end{equation}
This jump is in general non-zero because the value of $E'(y_j)$ follows from the bulk EoM and in general does not vanish.
There are no such jumps in 2-brane models because they involve only orbifold branes at which all fields must be continuous because of $\mathbb{Z}_2$ symmetry.

We are interested in the effective 4D description of the considered 5D models so we expand $Q$ in its KK modes:
\begin{equation}
\label{eq:Qn_def}
Q(x^\mu,y)=\sum_n Q_{m^2_n}(y) f_{m^2_n}(x^\mu)\,,
\end{equation}
where $f_{m^2}(x^\mu)$ are solutions of the 4D EoM:
$\Box_4 f_{m^2}(x^\mu) = m^2 f_{m^2}(x^\mu)$.
Using eq.~\eqref{eq:EoM_F} and the definition \eqref{eq:Q_def} one can find that the bulk equation which must be fulfilled by the profile of the KK mode with 4D masses squared equal $m^2$, $Q_{m^2}(y)$, may be cast in the form of the following Sturm-Liouville equation
\begin{equation}
\label{eq:EoM_Qn}
-(p\,Q_{m^2}')'+q\,Q_{m^2}=m^2\, p\,Q_{m^2}\,,
\end{equation}
where primes denote derivatives with respect to $y$ while $p$ and $q$ are background dependent functions: 
\begin{equation}
\label{eq:pq}
p:=\frac{3}{2\kappa^2}\frac{e^A}{(\phi')^2}\,,
\qquad\qquad q:=e^A\,.
\end{equation}

Of course the EoM \eqref{eq:EoM_Qn} is not enough to find the spectrum of the KK modes of $Q$. The corresponding junction conditions are also necessary. The first set of such JCs may be obtained from the continuity of the GW field $\Phi$ and its perturbation $F_3$. Expressing $F_3$ in terms of $Q_{m^2}$ (with the help of eqs.~\eqref{eq:F3F2F1}, \eqref{eq:F2F1}, \eqref{eq:Q_def} and \eqref{eq:Qn_def}) and using the bulk EoM \eqref{eq:EoM_A''} and the JCs \eqref{eq:JC_A'}, \eqref{eq:JC_phi'} and \eqref{eq:E''} one finds
\begin{equation}
\label{eq:JC_Qn1}
\left[\frac{Q_{m^2}'}{\phi'}\right]_j-\frac{\kappa^2}{3}\,\frac{[\phi']_j}{[A']_j}\,[Q_{m^2}]_j = 0\,.
\end{equation}
Profiles $Q_{m^2}(y)$ of the KK modes may be discontinuous across branes similarly as the full 5D field $Q$ as shown in eq.~\eqref{eq:[Q]}.

Another set of JCs may be inferred from the EoM for the GW scalar field $\Phi$ (obtained by varying the action \eqref{eq:S} with respect to $\Phi$). Integration of such EoM over infinitesimally small interval $[y_j-\epsilon,y_j+\epsilon]$ containing the position of the $j$-th brane leads, in the first order in small scalar perturbations, to the following equation:
\begin{equation}
\label{eq:JC123}
\left[\phi'\right]_j F_2(y_j) -\left[F_3'\right]_j + e^{-A(y_j)} F_3(y_j)U_j^{\prime\prime}(\phi(y_j))=0\,.
\end{equation}
We want to express the above condition in terms of the profiles $Q_{m^2}$. It follows from eqs.~\eqref{eq:F3F2F1}, \eqref{eq:Q_def} and \eqref{eq:Qn_def} that the second term on the l.h.s.~of \eqref{eq:JC123} leads to the appearance of the second derivative of $Q_{m^2_n}$. After eliminating $Q^{\prime\prime}_{m^2}$ with the help of the bulk EoM \eqref{eq:EoM_Qn}, equation \eqref{eq:JC123} may be rewritten as the following condition
\begin{equation}
\label{eq:JC_Qn2}
B_j\left(\left<\frac{Q_{m^2}'}{\phi'}\right>_j-\left[\frac{Q_{m^2}'}{\phi'}\right]_j
\frac{\left<\phi'\right>_j}{\left[\phi'\right]_j}\right)
+m^2 \left[\frac{Q_{m^2}}{\phi'}\right]_j = 0\,,
\end{equation}
where the angle bracket denotes averaging over both sides of a given brane: $\left<X\right>_j:=\frac12\left(X(y_j^-)+X(y_j^+)\right)$. The coefficients $B_j$ depend on the background solution and are given by
\begin{equation}
\label{eq:B_def}
B_j:=e^{-A(y_j)}\,U^{\prime\prime}_j\big(\phi(y_j)\big)
-\left[A'\right]_j
-\left[\frac{\phi^{\prime\prime}}{\phi'}\right]_j\,.
\end{equation}
The JCs \eqref{eq:JC_Qn1} and \eqref{eq:JC_Qn2} must be fulfilled at each brane. However, they simplify considerably for the two orbifold branes which are located at the $\mathbb{Z}_2$ fixed points. 
All considered scalar fields are $\mathbb{Z}_2$ even so their derivatives with respect to $y$ are $\mathbb{Z}_2$ odd. Using this simple observation one can easily show that at both orbifold branes condition \eqref{eq:JC_Qn1} is automatically fulfilled while condition \eqref{eq:JC_Qn2} simplifies to
\begin{align}
\label{eq:BC_Qn2_orbi_1}
\frac12B_1 Q_{m^2}'(y_1^+)+m^2  Q_{m^2}(y_1^+)&=0\,,
\\
\frac12B_2 Q_{m^2}'(y_2^-)-m^2  Q_{m^2}(y_2^-)&=0\,.
\label{eq:BC_Qn2_orbi_2}
\end{align}
This is one of the reasons for which analyses of 2-brane models are much simpler than those for multibrane ones considered in this work. At each brane in the former case one has to take into account only one simple condition, \eqref{eq:BC_Qn2_orbi_1} or \eqref{eq:BC_Qn2_orbi_2}, while in the later case two more complicated conditions, \eqref{eq:JC_Qn1} and \eqref{eq:JC_Qn2}, must be analyzed.

Now we have all ingredients which may be used to find the spectrum of the KK modes of the scalar perturbation $Q$. There is a KK mode with the mass squared equal $m^2$ if the corresponding solution of the bulk EoM \eqref{eq:EoM_Qn} fulfills BCs at both orbifold branes, \eqref{eq:BC_Qn2_orbi_1} and \eqref{eq:BC_Qn2_orbi_2}, and JCs 
\eqref{eq:JC_Qn1} and \eqref{eq:JC_Qn2} at all intermediate branes.
Of course we are not going to calculate the full spectrum of the scalar KK modes (in most cases it is possible only to find some of the masses numerically). We want to investigate stability of multibrane models so it is enough to determine the sign of the smallest $m_n^2$. Any negative $m_n^2$ indicates instability of a given model.

%%%%%%%%%%%%%%%%%%%%%%%%%%%%%%%%%%%%%%%%%%%%%%%%%%%%%%%%%%%%%%%%%
\subsection{How many radions are there?}
%%%%%%%%%%%%%%%%%%%%%%%%%%%%%%%%%%%%%%%%%%%%%%%%%%%%%%%%%%%%%%%%%

There is some discussion in the literature on the number of radions in mulitibrane 5D models (see e.g.~\cite{Lee:2021wau,Cai:2022geu,Girmohanta:2023sjv}). In any such discussion one should first of all adopt a precise definition of radions because it is not possible to count objects without their clear definition.

The situation is simple in the case of the original RS model (see e.g.~\cite{Kofman:2004tk}). There is one massless KK mode of 5D scalar perturbations of the metric. All massive scalar KK modes are absorbed by massive tensor KK modes i.e.~by massive gravitons. This massless 4D scalar is usually called graviscalar or radion (the name ``radion'' is more commonly used).  Vanishing of its mass is related to the fact that the positions of the branes in the RS model are not fixed. Background with arbitrary interbrane distance is a solution of the bulk EoMs and brane BCs.

Situation complicates a little bit when the 5D GW scalar field is added. Perturbations of such field give rise to an infinite tower of 4D scalar KK modes. Usually only the lightest of these modes is considered as the 
radion, at least in 2-brane models. Some authors argue there are two radions in the case of 3-brane models. Following such attitude, should one talk about $n$ radions in models with $n$ (independent) interbrane distances?

As we have shown in this Section, there is just one infinite tower of scalar KK modes irrespective of the number of branes. 
It seems that the most appropriate way is to name all of these KK modes as radions.
%\footnote{The name ``radions'' may be to some extend misleading because in all models with GW mechanism there are infinitely many radions while there is only a finite number of radii (interbrane distances).}. 
It is true that in a model with $N$ branes there are $N-1$ 
(independent) interbrane distances but, as will be discussed later, all 4D scalar KK modes are somehow related to those distances.
Limiting the name ``radions'' to only some of the scalar KK modes, with their number depending on the number of branes, does not seem very well justified. Maybe instead of asking about the number of such ``radions'' one should ask about the number of light radions. Then of course one has to choose a scale separating light from heavy states.

There are several analogies between defined here radions and gravitons (some of them will be discussed later).
There is one infinite tower of 4D gravitons (tensor perturbations) and 
one infinite tower of 4D radions (scalar perturbations). 
A very important difference between gravitons and radions is that by construction the lightest graviton is massless while 
the smallest radion mass squared may be negative, zero or positive.
The main goal of this work is to find criteria which allow to check which of these possibilities is realized in a given multibrane model.

%%%%%%%%%%%%%%%%%%%%%%%%%%%%%%%%%%%%%%%%%%%%%%%%%%%%%%%%%%%%%%%%
\section{Conditions for background stability}
\label{sec:stability}
%%%%%%%%%%%%%%%%%%%%%%%%%%%%%%%%%%%%%%%%%%%%%%%%%%%%%%%%%%%%%%%%

Background solutions of multibrane models were discussed in Section \ref{sec:5Dmodels}. In Section \ref{sec:perturbations} we presented bulk EoMs and brane JCs for profiles along the 5th dimension of the corresponding radions i.e.~scalar perturbations around those backgrounds. Solving such EoMs and JCs one may find the spectrum of radions. The smallest radion mass squared is related to stability. A given background is stable (unstable) if this mass squared is positive (negative). Let us first investigate the borderline case of a massless radion.

%%%%%%%%%%%%%%%%%%%%%%%%%%%%%%%%%%%%%%
\subsection{Massless radions}
\label{sec:m=0}
%%%%%%%%%%%%%%%%%%%%%%%%%%%%%%%%%%%%%%

It is not difficult to check that the solution of the bulk EoM \eqref{eq:EoM_Qn} between branes $j$ and $j+1$ with $m=0$ may be written in the following form
\begin{align}
\label{eq:Q0}
Q_0(y)&=c_1e^{-2A(y)}+c_2A'(y)\,e^{A(y)}\int_{y_j}^y e^{-3A(y')} \d y',
\\
\label{eq:Q0'}
Q'_0(y)&=(c_2-2c_1)A'(y)\,e^{-2A(y)}
+c_2\frac{\kappa^2}{3}\big(\phi'(y)\big)^2e^{A(y)}\int_{y_j}^y e^{-3A(y')} \d y'\,.
\end{align}
Now we try to construct the full (not only in one interbrane section of the bulk) profile of a massless radion. Let us start at the first orbifold brane. The BC \eqref{eq:BC_Qn2_orbi_1} in the case of $m^2=0$ and $B_1\ne0$ reduces just to a simply equation: $Q'(y_1^+)=0$. Thus, from eqs.~\eqref{eq:Q0} and \eqref{eq:Q0'} it follows that the massless radion profile between the first orbifold brane and the first intermediate brane is given by
\begin{align}
\label{eq:Q0y1}
Q_0(y)&=e^{-2A(y)}+2A'(y)e^{A(y)}\int_{y_1}^y e^{-3A(y')} \d y'\,,
\\
\label{eq:Q0'y1}
Q'_0(y)&=\frac23\kappa^2\big(\phi'(y)\big)^2e^{A(y)}\int_{y_1}^y e^{-3A(y')} \d y'\,,
\end{align}
(where we fixed the normalization by choosing $c_1=1$). These formulae give as values of $Q_0(y_{I_1}^-)$ and $Q'_0(y_{I_1}^-)$ just ``before'' the brane located at $y_{I_1}$. Values of $Q_0(y_{I_1}^+)$ and $Q'_0(y_{I_1}^+)$
just ``after'' that brane may be calculated with the help of the JCs \eqref{eq:JC_Qn1} and \eqref{eq:JC_Qn2}. The latter in the case of $m^2=0$
and $B_{j}\ne0$ reads
\begin{equation}
\left<\frac{Q_0'}{\phi'}\right>_{j}-\left[\frac{Q_0'}{\phi'}\right]_{j}
\frac{\left<\phi'\right>_{j}}{\left[\phi'\right]_{j}}=0\,,
\end{equation}
which may be rewritten in the form
\begin{equation}
\frac{1}{\left[\phi'\right]_{j}}
\left(Q_0'(y_j^-)\frac{\phi'(y_j^+)}{\phi'(y_j^-)}
-Q_0'(y_j^+)\frac{\phi'(y_j^-)}{\phi'(y_j^+)}
\right)=0\,.
\end{equation}
For $\phi'(y_j^\pm)\ne0$ this equation simplifies to 
\begin{equation}
\left[\frac{Q_0'}{(\phi')^2}\right]_{j}=0\,.
\end{equation}
It follows from this condition with $j=I_1$ that the expression \eqref{eq:Q0y1} for $Q_0'$ is valid also for $y=y_{I_1}^+$.
The radion profile $Q_0$ must fulfill also the second JC \eqref{eq:JC_Qn1}. Using background JCs \eqref{eq:JC_A'} and \eqref{eq:JC_phi'}, condition \eqref{eq:JC_Qn1} may be written in the form
\begin{equation}
\label{eq:[Q0]}
\left[Q_0\right]_j
=
\frac{3}{\kappa^2}\left[\frac{Q_0'}{\phi'}\right]_j\frac{\left[A'\right]_j}{\left[\phi'\right]_j}\,.
\end{equation} 
As we already shown, eq.~\eqref{eq:Q0'y1} is valid at both sides of the brane so it may be used to calculate the jump of $Q_0'/\phi'$:
\begin{equation}
\left[\frac{Q_0'}{\phi'}\right]_{I_1}
=
\frac23\kappa^2\left[\phi'\right]_{I_1}e^{A(y_{I_1})}\int_{y_1}^{y_{I_1}} e^{-3A(y')} \d y'\,.
\end{equation}
Substituting this into \eqref{eq:[Q0]} (with $j=I_1$) we obtain
\begin{equation}
\left[Q_0\right]_{I_1}=2\left[A'\right]_{I_1}e^{A(y_{I_1})}\int_{y_1}^{y_{I_1}} e^{-3A(y')} \d y'\,,
\end{equation}
which shows that expression \eqref{eq:Q0y1} also is valid for $y=y_{I_1}^+$.
Thus, formulae \eqref{eq:Q0y1} and \eqref{eq:Q0'y1} are both valid in two interbrane sections between the first orbifold brane and the second intermediate brane. The same steps may be repeated at all intermediate branes leading to the conclusion that $Q_0$ and $Q_0'$ given by \eqref{eq:Q0y1} and \eqref{eq:Q0'y1} fulfill the EoM in the whole bulk, BC at the first orbifold brane and JCs at all intermediate branes. The last condition to be checked is the BC at the second orbifold brane \eqref{eq:BC_Qn2_orbi_2} which for $m=0$ with the help of \eqref{eq:Q0'y1} may be written as
\begin{equation}
B_2\,\phi'(y_2^-)=0\,.
\end{equation}
There is a massless radion if the above equality if satisfied. 
Repeating the above procedure of solving the EoM and JCs but starting from the second orbifold brane one arrives at the conclusion that there is a massless radion if 
\begin{equation}
B_1\,\phi'(y_1^+)=0\,.
\end{equation}
The combination of the last two conditions is know to be the necessary and sufficient condition for the existence of a massless radion in 2-brane models \cite{Lesgourgues:2003mi}. 
We have shown that vanishing of any of the quantities: $B_1$, $B_2$, $\phi'(y_1)$ or $\phi'(y_2)$ is also a sufficient condition in the case of multibrane models. However, it is no longer a necessary condition. In our construction of the massless radion profiles so far we assumed that at all intermediate branes $B_{I_i}\ne0$ and $\phi'(y_{I_i}^\pm)\ne0$. Let us now consider situation when at one of the intermediate branes one of these quantities does vanish. In such a case one may construct two solutions with $m=0$: one starting from $y_1$ and second starting from $y_2$. The first is valid for $y<y_I$ and the second for $y>y_I$ where $y_I$ is the position of that brane at which $B_I\,\phi'(y_I^-)\,\phi'(y_I^+)=0$. We have to check whether both JCs at that brane may be fulfilled. This is very easy when $B_I=0$. In such a case JC \eqref{eq:JC_Qn2} with $m=0$ is automatically satisfied while JC \eqref{eq:JC_Qn1} fixes the relative normalization of 
the two parts of the solution, one to the left and one to the right of the considered brane. The situation is somewhat more complicated in the case of vanishing $\phi'$. Without loss of generality we may choose $\phi'(y_I^-)=0$. Expression \eqref{eq:Q0'y1} tells us that in such a case the ratio $Q'_0(y_I^-)/\phi'(y_I^-)=0$. It is easy to check that JC \eqref{eq:JC_Qn2} is then automatically satisfied irrespective of the values of $Q_0'$ and $\phi'(y_I^+)$. The remaining JC \eqref{eq:JC_Qn1} is well defined (it follows from \eqref{eq:Q0'y1} that $Q_0'/\phi'$ is finite even for vanishing $\phi'$) and again determines the relative normalization of the two parts of the solution.

The above discussion shows that there is a massless radion if at any brane, orbifold or intermediate one, at least one of quantities, $B_j$, $\phi'(y_j^-)$ or $\phi'(y_j^+)$ vanishes. On the other hand, if all these quantities at all branes are non-zero the radion EoM and JCs with $m=0$ have no solutions. Thus the necessary and sufficient condition for the existence of a massless radion is
\begin{equation}
\label{eq:condition_for_m=0}
\prod_j B_j\phi'(y_j^-)\phi'(y_j^+)=0\,,
\end{equation}
where the product is taken over all branes.

There is another way to see that the above equality is the sufficient condition for the existence of a massless radion. It is possible to show that the bulk EoM \eqref{eq:EoM_Qn} with JCs \eqref{eq:JC_Qn1}-\eqref{eq:BC_Qn2_orbi_2} correspond to some variational problem. Multiplying the bulk EoM by $Q(y)$, integrating over the full interval $[y_1,y_2]$ and using all JCs one may express the eigenvalue $m^2_n$ in terms of the corresponding eigenfunction $Q_{m^2_n}$ as follows:
\begin{equation}
\label{eq:m2n}
m_n^2
=
\frac{\int p(Q_{m^2_n}')^2+\int qQ_{m^2_n}^2\,+\sum_{I_i} q\frac{[Q_{m^2_n}]_{I_i}^2}{[A']_{I_i}}}
{\int pQ_{m^2_n}^2
+\left.\frac{2}{B_1}pQ_{m^2_n}^2\right|_{y_1^+}
+\left.\frac{2}{B_2}pQ_{m^2_n}^2\right|_{y_2^-}
+\sum_{I_i} \,\frac{3e^A}{2\kappa^2}\frac{1}{B_{I_i}}\left[\frac{Q_{m^2_n}}{\phi'}\right]_{I_i}^2}\,.
\end{equation}
Without intermediate branes the last terms in the denominator and numerator are absent and the expression reduces to the analogous one found for 2-brane models \cite{Lesgourgues:2003mi}.
The integrals in the above formula should be calculated over the sum of all interbrane segments: 
$\int\equiv\int_{y_1}^{y_{I_1}}+\int_{y_{I_1}}^{y_{I_2}}+\ldots+\int_{y_{I_{N-2}}}^{y_2}$.
The smallest eigenvalue, $m_{\rm min}^2$, can be obtained by minimizing the r.h.s.~of the above equation over all functions Q defined on the interval
$[y_1,y_2]$ which are smooth over each interbrane section but may have discontinuities at intermediate branes:
\begin{equation}
\label{eq:m2min}
m_{\rm min}^2
=
\min_Q
\left(
\frac{\int p(Q')^2+\int qQ^2\,+\sum_{I_i} q\frac{[Q]_{I_i}^2}{[A']_{I_i}}}
{\int pQ^2
+\left.\frac{2}{B_1}pQ^2\right|_{y_1^+}
+\left.\frac{2}{B_2}pQ^2\right|_{y_2^-}
+\sum_{I_i} \,\frac{3e^A}{2\kappa^2}\frac{1}{B_{I_i}}\left[\frac{Q}{\phi'}\right]_{I_i}^2}\right)\,.
\end{equation}
%\end{widetext}
%
In this formula the profile functions $Q$ and their derivatives $Q'$ appear always squared. The background-dependent functions $p$ and $q$ are non-negative.
The only terms which could be negative are brane parameters $B_j$ and the jumps $[A']_{I_i}$ at the intermediate branes. Such possibilities will be discussed later. Now we assume $[A']_{I_i}>0$ and $B_j\ge0$ from which it follows that the expression in the bracket is non-negative for arbitrary $Q$. Moreover, that expression approaches zero from above if any $B_j\to0^+$ or if $\phi'\to0$ at any brane (in the case of the orbifold branes $(\phi')^{-2}$ is ``hidden'' in $p$). The lower limit of positive numbers which may be arbitrary small is of course equal zero. This is in full agreement with the condition \eqref{eq:condition_for_m=0}.

In models for which equation \eqref{eq:condition_for_m=0} is satisfied there is at least one massless radion. However, in some models more such radions may be present if more factors in \eqref{eq:condition_for_m=0} vanish. Let us first discuss this for a simple 2-brane model. The bulk EoM \eqref{eq:EoM_Qn} with $m^2=0$ has two linearly independent solutions given by \eqref{eq:Q0} with arbitrary values of $c_1$ and $c_2$. If both $B_1$ and $B_2$ are equal zero than both BCs \eqref{eq:BC_Qn2_orbi_1} and  \eqref{eq:BC_Qn2_orbi_2} with $m=0$ are automatically satisfied. So, each of the independent bulk solutions corresponds to a different 4D KK mode with $m=0$. It is obvious that having  $B_1=B_2=0$ is the only way to have two massless radions in a 2-brane model. If at least one $B_i$ is non-zero the corresponding BC, \eqref{eq:BC_Qn2_orbi_1} or \eqref{eq:BC_Qn2_orbi_2}, eliminates one of the solutions of the bulk EoM \eqref{eq:EoM_Qn}. This reasoning may be easily generalized for multibrane models by applying the above arguments for each interbrane section of the bulk. The result is that in a model in which $n_B>0$ parameters $B_j$ are equal zero there are $2^{n_B-1}$ massless radions. For $n_B=0$ there is just one massless radion if (and only if) condition \eqref{eq:condition_for_m=0} is fulfilled.

%%%%%%%%%%%%%%%%%%%%%%%%%%%%%%%%%%%%%%
\subsection{Massive radions}
\label{sec:m>0}
%%%%%%%%%%%%%%%%%%%%%%%%%%%%%%%%%%%%%%

Equation \eqref{eq:m2n} shows that masses squared of all radions are positive if all brane parameters $B_j$ and jumps of $A'$ at all intermediate branes are positive and if $\phi'$ is non-zero everywhere. The statement about $\phi'$ needs some explanation. 
One could argue that it is too strong because in the previous Subsection it was shown that at least one radion is massless if $\phi'$ has limit equal 0 for $y$ approaching any of the brane positions and not for any $y$ in the bulk. If all such limits are non-vanishing, and in addition conditions \eqref{eq:Bj>0} and \eqref{eq:Uj>0} are satisfied,  than all boundary terms in formula \eqref{eq:m2n} are positive. However, this occurs to be not necessarily sufficient for the positivity of all radions masses squared $m_n^2$. The reason is that $p$ defined by \eqref{eq:pq} is proportional to $(\phi')^{-2}$ so it diverges at points in the bulk at which $\phi'=0$ leading to problems with the integrals in both the numerator and denominator in the expression \eqref{eq:m2n}. Situation with $\phi'$ vanishing somewhere in the bulk requires more careful treatment and will be discussed in the next Subsection. 

Special attention must be paid also when at any intermediate brane $[A']_I=0$ because in such cases the last term in the numerator of eq.~\eqref{eq:m2min} seems to be ill-defined. This should be analyzed as the limit of $[A']_I\to0$. It follows from the JC \eqref{eq:JC_Qn1} that in such a case $[Q]_I\to0$ with the ratio $[Q]_I/[A']_I$ approaching a finite value. Thus, $[Q]^2_I/[A']_I\to0$ and the last term in the numerator of \eqref{eq:m2min} vanishes in the limit of $[A']_I\to0$. If there is no other intermediate brane with negative tension than there are no negative contributions in the numerator of \eqref{eq:m2min}. The minimum of a set of non-negative numbers can not be negative, so there are no tachyonic radions when $[A']_I=0$ (with all $B_j>0$ and $\phi'\ne0$). In principle there could be a massless radion in such situation, but it is not the case. It is easy to check the JCs \eqref{eq:JC_Qn1} and \eqref{eq:JC_Qn2} at brane $I$ can not be simultaneously fulfilled for the radion profile \eqref{eq:Q0} when $[A']_I=0$.

It follows from the above arguments that masses squared of all radions are positive if the following three conditions are simultaneously satisfied:
\begin{enumerate}
\item
All brane parameters $B_j$ are positive:
\begin{equation}
\label{eq:Bj>0}
B_j=e^{-A(y_j)}U^{\prime\prime}_j
\big(\phi(y_j)\big)-\left[A'\right]_j-\left[\frac{\phi^{\prime\prime}}{\phi'}\right]_j>0\,.
\end{equation}
\item
Derivative of the GW scalar background is nowhere vanishing:
\begin{equation}
\label{eq:phi'<>0}
\begin{cases}
\hspace{42pt}
\phi'(y)\ne0 \qquad \text{in the bulk} 
\\
\lim_{y\to y_j^\pm}\phi'(y)\ne0 \qquad \text{at all branes} 
\end{cases}\,.
\end{equation}
\item
All intermediate branes $I_i$ have non-negative tensions from which it follows that all jumps of $A'$ are also non-negative:
\begin{equation}
\label{eq:Uj>0}
\left[A'\right]_{I_i} \ge0 \,.
\end{equation}
\end{enumerate}
These conditions are sufficient for the positivity of all $m_n^2$ so for the stability of a considered multibrane configuration.

%%%%%%%%%%%%%%%%%%%%%%%%%%%%%%%%%%%%%%
\subsection{Tachyonic radions}
\label{sec:m<0}
%%%%%%%%%%%%%%%%%%%%%%%%%%%%%%%%%%%%%%

Conditions \eqref{eq:Bj>0}-\eqref{eq:Uj>0} are sufficient for all radion masses squared to be positive (which guaranties that a considered multibrane background configuration is stable).
Now we will prove that they are also the necessary ones. 
In order to do this we will show that violation of any of these conditions results in at least one massless or tachyonic radion. Massless radions were already discussed in Section \ref{sec:m=0} where it was shown that (at least one) such radion exists if condition \eqref{eq:condition_for_m=0} is fulfilled. So, now we will investigate models when any of the conditions 
\eqref{eq:Bj>0}-\eqref{eq:Uj>0} is violated ``stronger'' than just by \eqref{eq:condition_for_m=0}.

\subsubsection*{\boldmath$B_j<0$}

Analysis of condition \eqref{eq:Bj>0} is quite simple.   
It is enough to use the following obvious property of variational calculus.
Let us consider two problems of minimization of given functionals. If for
every function the functional in the first problem is not bigger than the
functional in the second problem then the first minimum can not be bigger
than the second one.
We apply this to two problems of minimizing the expression in the r.h.s.~of \eqref{eq:m2min} for a given background (the same $A(y)$ and $\phi(y)$ so also the same positive $p(y)$ and $q(y)$) but with different values of one of the parameters $B_j$.\footnote{It is possible to have different values of $B_j$ for the same background because $B_j$ contains the second derivative of the corresponding brane potential, $U_j^{\prime\prime}$, which does not appear in the background EoMs \eqref{eq:EoM_A''}--\eqref{eq:EoM_A'phi'} and JCs \eqref{eq:JC_phi'}--\eqref{eq:BC_A'_12}.}
 Let the value of one $B_j$ be bigger in the second problem.
Bigger $B_j$ means smaller (at least not bigger) denominator in \eqref{eq:m2min} and thus bigger (at least not smaller) the whole expression to be minimized. Applying the above mentioned rule of the variational calculus, we see that the lowest eigenvalue in the problem with bigger $B_j$ is bigger (at least not smaller) than the lowest eigenvalue in the problem with smaller $B_j$. The lowest eigenvalue, $m_{\rm min}^2$, is a monotonic, non–decreasing function of $B_j$.

We apply the above reasoning to a negative $B_j$. The monotonic character of   $m_{\rm min}^2$ as a function of $B_j$ means that the lowest eigenvalue for any negative $B_j$ can not be bigger than the one for vanishing $B_j$, which
was previously shown to be zero. So, we get $m_{\rm min}^2\le0$ for any $B_j<0$. On the other hand, $m_{\rm min}^2$ can not be equal to zero, because we have shown that a zero mode exists only if one of the $B_j$ vanishes (we consider a situation with conditions  \eqref{eq:phi'<>0} and \eqref{eq:Uj>0} satisfied, otherwise there is a zero mode for arbitrary $B_j$). 
Thus, a background with a negative value of any of the brane parameters $B_j$ is unstable because at least one radion is tachyonic.

\subsubsection*{\boldmath$\phi'=0$ \bf at some point in the bulk}

The situation with a background with $\phi'$ vanishing somewhere in the bulk is more complicated. In such cases the bulk EoM for the scalar KK modes $\eqref{eq:EoM_Qn}$ can not be used because function $p$ is singular at points at which $\phi'=0$. It is necessary to use appropriate Mukhanov-Sasaki variable $v$. Its KK modes, $v_{m^2_n}$, are related to the used so far  modes $Q_{m^2_n}$ via the relation
\begin{equation}
\label{eq:Q-v}
Q_{m^2_n}=\frac{e^{-2A}(\phi')^2}{m_n^2 A'}\left(\frac{e^{3A/2}A'}{\phi'}v_{m^2_n}\right)'\,.
\end{equation}
Substitution of such $Q_{m^2_n}$ into its EoM \eqref{eq:EoM_Qn} leads, after using the background EoMs, to the equation of motion for the new variable $v_{m^2_n}$ which may be written in the form
\begin{equation}
v_{m^2_n}^{\prime\prime}
+\left(
m_n^2
-\frac{15}{4}\left(A'\right)^2
-\frac{19}{6}\kappa^2\left(\phi'\right)^2
+\frac29\kappa^4\frac{\left(\phi'\right)^4}{\left(A'\right)^2}\right.
-\left.\frac43\kappa^2e^{-2A}V'\frac{\phi'}{A'}
+e^{-2A}V^{\prime\prime}
\right)v_{m^2_n}=0\,.
\label{eq:EoM_vn}
\end{equation}
The above EoM is explicitly regular also at points for which $\phi'=0$. The solution of this EoM in the first interbrane section in the limit of large negative $m^2$ is approximated by
\begin{equation}
v_{-\infty}(y)
\approx
c_+\exp\left(\sqrt{-m^2}(y-y_1)\right)
+c_-\exp\left(-\sqrt{-m^2}(y-y_1)\right)\,.
\end{equation}
Using the relation \eqref{eq:Q-v}, we obtain expressions for corresponding $Q_{-\infty}(y)$ and $Q'_{-\infty}(y)$. The BC at the first orbifold brane \eqref{eq:BC_A'_12} at $y_1$ can be fulfilled if $|c_+/c_-|={\cal{O}}(1)$ in the leading order in $1/m^2$. Hence $c_+$ is not small compared to $c_-$, and away from the first brane the solution is dominated by the exponentially growing term:
\begin{align}
\label{eq:Q-infty}
Q_{-\infty}(y)
&\approx
-\frac{c_+ e^{-A/2}\phi'(y)}{\sqrt{-m^2}}\exp\left(\sqrt{-m^2}(y-y_1)\right),
\\
\label{eq:Q'-infty}
Q'_{-\infty}(y)
&\approx
-c_+ e^{-A/2}\phi'(y)\exp\left(\sqrt{-m^2}(y-y_1)\right).
\end{align}
In the leading order in $1/m^2$ the JCs \eqref{eq:JC_Qn1} and \eqref{eq:JC_Qn2} reduce to 
\begin{equation}
\left[\frac{Q'_{-\infty}}{\phi'}\right]_{I_1}\approx0
\,,\qquad\qquad
\left[\frac{Q_{-\infty}}{\phi'}\right]_{I_1}\approx0\,,
\end{equation} 
respectively. 
One can see that these JCs are fulfilled when $Q_{-\infty}(y)$ and $Q'_{-\infty}(y)$ are given by eqs.~\eqref{eq:Q-infty} and \eqref{eq:Q'-infty} also in the second interbrane section of the bulk. Using the same arguments at all remaining intermediate branes one can show that expressions \eqref{eq:Q-infty} and \eqref{eq:Q'-infty} describe $Q_{-\infty}(y)$ and $Q'_{-\infty}(y)$ in the whole bulk (to the leading order in $1/m^2$).

The important feature of this solution $Q_{-\infty}(y)$ is that it is proportional to $\phi'(y)$ so it vanishes (and changes sign) at the point at which $\phi'$ vanishes (and changes sign). This should be compared to the behavior of $Q_0(y)$. Equation \eqref{eq:Q0'y1} shows that $Q'_0$ is non-negative so $Q_0$ is a monotonically growing positive function. Solutions $Q_{m^2}$ of the bulk EoM change continuously with $m^2$ so there must be such negative $\widetilde{m}^2$ for which $Q_{\widetilde{m}^2}$ is positive everywhere except just one point at which it is equal zero. This point must be at the position of the second orbifold brane\footnote{$Q$ and $Q'$ can not vanish at the same point because otherwise, being the solution of the second order differential equation, $Q$ would be zero in the whole section of the bulk.} at which $Q_{\widetilde{m}^2}(y_2)=0$ and $Q'_{\widetilde{m}^2}(y_2)<0$.

The l.h.s.~of the BC \eqref{eq:BC_Qn2_orbi_2} is positive for $Q_0$ and negative for $Q_{\widetilde{m}^2}$. Thus, there must be at least one value of $m^2$ between 0 and $\widetilde{m}^2<0$ for which the BC at $y=y_2$ is fulfilled. So, there must be at least one tachyonic radion.

\subsubsection*{\bf\boldmath$[A']<0$ at any intermediate brane}

Finally we consider situations with at least one negative tension intermediate brane $I_i$ for which, according to eq.~\eqref{eq:JC_A'}, $\left[A'\right]_{I_i}<0$. Let us first discuss a model with just one intermediate brane for which other conditions sufficient 
for the stability, i.e.~\eqref{eq:Bj>0} and \eqref{eq:phi'<>0}, are satisfied. For any value of $m^2$ and for each of the two interbrane sections one can construct the solution of the bulk EoM \eqref{eq:EoM_Qn} satisfying appropriate BC, \eqref{eq:BC_Qn2_orbi_1} or \eqref{eq:BC_Qn2_orbi_2}, at the corresponding orbifold brane. Such solutions, $Q_{m^2L}(y)$ and $Q_{m^2R}(y)$, give values of $Q$ and $Q'$ at both sides of the intermediate brane i.e.~at $y=y_I^-$ and $y=y_I^+$. The spectrum of radions consists of such values of $m^2$ for which both JCs at $y=y_I$, \eqref{eq:JC_Qn1} and \eqref{eq:JC_Qn2}, are fulfilled. Using JC \eqref{eq:JC_Qn2} to determine the relative normalization of solutions in two sections of the bulk one can write the l.h.s.~of JC \eqref{eq:JC_Qn1} in the following form
\begin{equation}
\label{eq:Delta_1}
\Delta(m^2):=
\frac{Q_{m^2-}}{\phi'_-}\,
\frac{P_{m^2L}P_{m^2R}
+\frac{m^2}{B_I}\left(P_{m^2L}-P_{m^2R}\right)
+\frac{\kappa^2m^2\left[\phi'\right]_I^2}{3B_I\left[A'\right]_I}
+\frac{\kappa^2}{3\left[A'\right]_I}\left(P_{m^2R}(\phi'_-)^2-P_{m^2L}(\phi'_+)^2\right)}{P_{m^2R}\frac{\phi'_-}{\left[\phi'\right]_I}-\frac{m^2}{B_I}}\,,
\end{equation}
with the notation: $Q_{m^2-}:=Q_{m^2L}(y_I^-)$, $Q'_{m^2-}:=Q'_{m^2L}(y_I^-)$, $P_{m^2L}:=Q'_{m^2L}(y_I^{-})/Q_{m^2L}(y_I^{-})$, $P_{m^2R}:=Q'_{m^2R}(y_I^{+})/Q_{m^2R}(y_I^{+})$, $\phi'_\pm:=\phi'(y_I^\pm)$. 
Equation for $\Delta$ simplifies in the case of $m^2=0$:
\begin{equation}
\Delta(0)=
\frac{Q'_{0-}}{(\phi'_-)^2}\,\left[\phi'\right]_I
\left(
1+\frac{\kappa^2}{3\left[A'\right]_I}\left(\frac{(\phi'_-)^2}{P_{0L}}-\frac{(\phi'_+)^2}{P_{0R}}\right)
\right)\,.
\end{equation}
Bulk solutions with $m^2=0$ are known. In the left bulk section it is given by eqs.~\eqref{eq:Q0y1} and \eqref{eq:Q0'y1} while in the right section by the same equations but with $y_1$ replaced with $y_2$. Using such solutions one finds
\begin{equation}
\label{eq:Delta0}
\Delta(0)=
\frac{\kappa^2}{3}\,e^{-2A}\,\frac{\left[\phi'\right]_I}{\left[A'\right]_I}
\,\frac{\int_{y_1}^{y_2}e^{-3A}}{\int_{y_I}^{y_2}e^{-3A}}\,Q(0)\,.
\end{equation}

The expression \eqref{eq:Delta_1} for $m^2\ne0$ simplifies in the limit $B_I\to\infty$ (stiff potential on the intermediate brane):
\begin{equation}
\Delta(m^2)
\rightarrow
\frac{Q'_{m^2-}}{(\phi'_-)^2}\,\left[\phi'\right]_I
\left(
1+\frac{\kappa^2}{3\left[A'\right]_I}\left(\frac{(\phi'_-)^2}{P_{m^2L}}-\frac{(\phi'_+)^2}{P_{m^2R}}\right)
\right).
\end{equation}
Substituting solutions \eqref{eq:Q-infty} and \eqref{eq:Q'-infty} to the above expression one finds that for large negative $m^2$ in the leading order in $1/|m^2|$ 
\begin{equation}
\label{eq:Delta-infty}
\lim_{B_I\to\infty}
\Delta(m^2)
\approx
\sqrt{|m^2|}\,e^{A/2}
\,e^{\sqrt{|m^2|}(y_I-y_1)}
\,\frac{\left[\phi'\right]_I}{\phi'_-\phi'(0)}Q(0)\,.
\end{equation}
Comparing \eqref{eq:Delta0} and \eqref{eq:Delta-infty} one can see that the relative sign of the l.h.s.~of the JC \eqref{eq:JC_Qn1} in the cases of $m^2=0$ and large negative $m^2$ is the same as the sign of $\left[A'\right]_I$ ($\phi'_-\phi'(0)$ is positive because of the condition \eqref{eq:phi'<>0}). 
The l.h.s.~of the JC \eqref{eq:JC_Qn1} changes continuously with $m^2$ so in models with $[A']_I<0$ there must be at least one negative $m^2$ for which JC \eqref{eq:JC_Qn1} is fulfilled.

So far we have shown that in models with just one intermediate brane at least one radion is tachyonic if the tension of that brane is negative and the corresponding brane parameter $B_I$ is large enough. It is easy to generalize this result to all models with arbitrary number of intermediate branes and arbitrary (positive\footnote{If any $B_{I_i}$ is negative the background configuration is unstable anyway.}) values of brane parameters $B_{I_i}$ if any of those branes has negative tension. Additional branes present between orbifold branes and the considered brane with negative tension do not change the arguments given for the 3-brane system because (as was shown in Section \ref{sec:m=0} and in the discussed above case with $\phi'=0$) solutions with $m^2=0$ and the leading contribution for large negative $m^2$ have the same functional forms (i.e.~\eqref{eq:Q0} and \eqref{eq:Q-infty}, respectively) on both sides of a given intermediate brane. Thus, respective solutions satisfying boundary conditions at each orbifold brane may be extended to both sides of the negative tension brane in question.

Of course the spectrum of radions does depend on the values of the brane parameters $B_j$, also for models with some $[A']_I<0$. However, if there is at least one tachyonic radion for one set of positive values of parameters $B_j$ there must be at least one tachyonic radion for arbitrary values of $B_j$. The reason is as follows. The profile of a tachyonic radion $Q_t(y)$ satisfies eq.~\eqref{eq:m2n} with some fixed positive values of $B_j$ and negative $m^2$. Thus the numerator of \eqref{eq:m2n} with $Q=Q_t$ must be negative (because the denominator is positive). The minimal $m^2$ for different set of values of $B_j$ is given by the expression \eqref{eq:m2min} with appropriate values of $B_j$. The r.h.s.~of that expression is negative for $Q=Q_t$ so also the minimum must be negative. 

The discussion above eq.~\eqref{eq:Bj>0} shows that intermediate branes for which $[A']_I=0$ do not contribute to the numerator of the expression \eqref{eq:Uj>0} so the presence of such branes can not lead to any tachyonic radions. 
Thus, we have proven that $[A']_{I_i}\ge0$ is a necessary condition for the stability of a given background brane configuration because $[A']_I<0$ at any intermediate brane results in at least one tachyonic radion.

%%%%%%%%%%%%%%%%%%%%%%%%%%%%%%%%%%%%%%%%%%%%%%%%%%%%%%%%%%%%%%%%%%%%%
\section{Some properties of radions}
\label{sec:properties}
%%%%%%%%%%%%%%%%%%%%%%%%%%%%%%%%%%%%%%%%%%%%%%%%%%%%%%%%%%%%%%%%%%%%%

The smallest radion mass squared may be negative, vanishing or positive. 
The conditions necessary and sufficient to realize each of these possibilities in a given multibrane model were presented in the previous Section and constitute the main result of this work. Let us now discuss in addition some properties of radions which are sometimes considered in the literature.

%%%%%%%%%%%%%%%%%%%%%%%%%%
\subsection{Light radions}
%%%%%%%%%%%%%%%%%%%%%%%%%%

We adopt the definition of radions as 4D KK modes of 5D scalar perturbations around a given 5D background solution of the appropriate Einstein equations.
So, strictly speaking, the number or radions is infinite, similarly as the number of all, massless and massive, gravitons in the same model. More interesting and nontrivial question is that about the number of light radions\footnote{
This is another similarity to gravitons. For example, models with very light first massive graviton seems to be very interesting, see e.g.~\cite{Kogan:1999wc,Kogan:2000vb}.}.
Some approximate results may be obtained by analytical methods but usually in the limits of negligible back-reaction (of the GW scalar on the metric) and with the stiff potential approximation. Such approximation simplifies calculations but does not seem very natural. In our notation it corresponds to sending all brane parameters $B_j$ to infinity. As we have shown, this limit results in maximal possible radion masses so is not very well suited for the analysis of light radions.

Unfortunately, in most cases numerical calculations are necessary to find radion masses. Moreover, high precision calculations are essential in models addressing the hierarchy problem. However, using only the analytical approach adopted in this work one can find the possible number of light radions. It was shown in Section \ref{sec:m=0} that the number of strictly massless radions is related to the number of vanishing brane parameters $B_j$. The radion masses (eigenvalues of appropriate Sturm-Liouville problem) should changed continuously with the parameters so the number of (very) light radions should be related in the same way to the number of (very) small (but positive) parameters $B_j$. It is worth noticing that values of these parameters are not fully determined by the background solution i.e.~by fixing positions of all branes, the functions $A(y)$ and $\phi(y)$ and their derivatives in the bulk and their jumps at the branes. Parameters $B_j$ defined in \eqref{eq:B_def} depend also on the second derivatives of the brane potentials. Thus, it is possible to have exactly the same background but different brane parameters so different radion masses.

One radion may be very light even when none of the parameters $B_j$ is small. This follows from another possibility of having a strictly massless radion. One radion has vanishing mass if $\phi'$ has vanishing limit for $y$ approaching the position of any of the branes. When one of such limits is non-zero but very small one of the radions becomes very light. Yet another possibility is very small but non-zero $\phi'$ somewhere in the bulk. I leave more detailed analysis of light radion for a future work.

%%%%%%%%%%%%%%%%%%%%%%%%%%
\subsection{Radions vanishing in some sections of the bulk}
%%%%%%%%%%%%%%%%%%%%%%%%%%

Quite often in the literature radion(s) are discussed in connection to distance(s) between branes. In the case of multibrane models a given radion is usually considered to be related mostly to one of the interbrane distances if its profile vanishes, at least in some approximation, in other sections of the bulk. So, it is interesting to check if and when such concentration of radion's profile is possible. In order to address such questions one should consider the radion profile $Q(y)$ which vanishes in (at least) one of the bulk sections. Without loosing generality we may assume that $Q(y)=0$ for $y<y_I$ but $Q(y)\ne0$ for $y>y_I$, where $y_I$ is the position of one the intermediate branes. Of course such profile must fulfill both JCs  \eqref{eq:JC_Qn1} and \eqref{eq:JC_Qn2}. It is easy to check that this is possible only if
\begin{equation}
\label{eq:m2_Q=0}
m^2=\frac{\kappa^2}{3}\,\frac{B_I}{[A']_I}\,\phi'(y_I^-)\phi(y_I^+)\,,
\end{equation}
i.e.~for just one value of $m^2$. The above condition is necessary but not sufficient. The non-zero part of the profile must satisfy equation
\begin{equation}
\frac{Q'(y_I^+)}{Q(y_I^+)}
=
\frac{\kappa^2}{3}\,\frac{[\phi']_I}{[A']_I}\,\phi(y_I^+)\,
\end{equation}
(or the analogous condition for $y_I^+$ replaced with $y_I^-$ if $Q(y)$ should vanish in the bulk section to the right of the considered brane). 
The above equation gives (up to unimportant overall normalization) the boundary condition for the Sturm-Liouville equation \eqref{eq:EoM_Qn} with $m^2$ given by \eqref{eq:m2_Q=0}. In general the solution of such equation does not fulfill the boundary condition at the appropriate orbifold brane. In other words, usually none of the radions has mass squared given by \eqref{eq:m2_Q=0} so none of them may have profile vanishing in any of bulk sections. It is easy to verify for example that a massless radion may not be localized in the discussed here sens (this conclusion may be avoided in some models with more than one massless radion).

The profile of any of the radions may vanish in some of the bulk sections only in exceptional situations requiring fine tuning of parameters. Condition \eqref{eq:m2_Q=0} shows that this for sure is not possible for light massive radions in the stiff potential approximation with $B_I\to\infty$.

%%%%%%%%%%%%%%%%%%%%%%%%%%
\subsection{Relations between radions and distances}
%%%%%%%%%%%%%%%%%%%%%%%%%%

It is possible to find in the literature descriptions of the brane models in which distances between branes are related to vacuum expectation values of radion fields. Such descriptions are inadequate for the approach adopted in the present work. Distances between branes are determined by solving appropriate Einstein equations with radions not yet considered. Radions describe 4D modes of perturbations around those solutions. Vacuum expectation values of radions do vanish (almost by definition because backgrounds are defined by solutions wich are not perturbed yet). Nevertheless, it is very interesting to investigate the relations between radions and distances. It will occur that also distances other than those between branes are involved in an interesting way.

Let us start with a distance between two point, $A$ and $B$, with the same 4D coordinates but different 5th one: $({x^\mu},y_A)$ and $({x^\mu},y_B)$, respectively. These may be two vis-a-vis points on two branes. The distance between them is given by
\begin{equation}
\label{eq:d_AB}
d_{AB}=\int_{y_A}^{y_B}\sqrt{g_{55}}\,\d y\,,
\end{equation}
where in our perturbed metric \eqref{eq:metric_ansatz_perturb} 
\begin{equation}
\sqrt{g_{55}}=\sqrt{e^{-2A(y)}\left(1+2F_2(x^\mu,y)\right)}
\approx e^{-A(y)}+e^{-A(y)}F_2(x^\mu,y)\,.
\end{equation}
The first term on the r.h.s.~gives the distance between the two considered points in the background solution while the second term is responsible for the change of that distance due to perturbations. Such change is given by
\begin{equation}
\label{eq:delta_dAB_F2}
\delta d_{AB} \approx \int_{y_A}^{y_B}e^{-A(y)}F_2(x^\mu,y)\,\d y\,.
\end{equation} 
We would like to express $\delta d_{AB}$ in terms of the radions profiles $Q_{m_n^2}(y)$. First one has to find the relation between the relevant metric perturbation, $F_2(x^\mu,y)$, and the 5D scalar perturbation $Q(x^\mu,y)$. Using \eqref{eq:F2F1} and \eqref{eq:Q_def} one gets (arguments of the 5D fields will be suppressed to simplify the notation):
\begin{equation}
\label{eq:F2-QE}
F_2=-2Qe^{2A}+\frac12E^{\prime\prime}-\frac12A'E'\,.
\end{equation}
Substitution of \eqref{eq:F2-QE} to \eqref{eq:delta_dAB_F2} gives
\begin{equation}
\delta d_{AB} \approx 
-2 \int_{y_A}^{y_B} e^A Q\,\d y 
+
\frac12 \int_{y_A}^{y_B}\left(e^{-A}E'\right)'\,\d y
=
-2 \int_{y_A}^{y_B} e^A Q\,\d y 
+
\frac12 \left(e^{-A} E'\right)\Big|_A^B\,.
\end{equation}
If $A$ and $B$ are two vis-a-vis points on two adjacent branes one may use relation \eqref{eq:[Q]} to write the change of the distance between these points in the following form
\begin{equation}
\delta d_{AB} 
\approx 
-2 \int_{y_A}^{y_B} e^A Q\,\d y 
+
\left.e^{A}\right|_{y_B}\,\frac{[Q]_B}{[A']_B}
-
\left.e^{A}\right|_{y_A}\,\frac{[Q]_A}{[A']_A}\,,
\end{equation}
where $[\ldots]_{A(B)}$ denotes jumps of some quantities at the brane located at $y=y_A(y_B)$. Finally one may use expansion \eqref{eq:Qn_def} to express the result in terms of the radion profiles
\begin{equation}
\delta d_{AB} 
\approx 
\sum_n f_{m_n^2}(t,\vec{x})
\left(
- 2 \int_{y_A}^{y_B} e^{A(y)} Q_{m_n^2}(y)\,\d y 
+e^{A(y_B)}\,\frac{[Q_{m_n^2}]_B}{[A']_B}
-
e^{A(y_A)}\,\frac{[Q_{m_n^2}]_A}{[A']_A}
\right).
\label{eq:delta_dAB-Qn}
\end{equation}
The function $f_{m_n^2}(t,\vec{x})$ describes the propagation of a radion with mass $m_n$ in the 4D space-time. 

Several comments are in order:
\begin{itemize}
\item[a)]
Propagating radions change distances between points on different branes but not the average distances between the branes. Functions $f_{m_n^2}(t,\vec{x})$ describe 4D plane waves so the changes of distances are zero when averaged over the corresponding wavelengths or periods. Propagating radions may be interpreted as waves of distortion/bending of branes without change of the average branes positions.
\item[b)]
In general propagation of any radion influences distances between points on different branes.

The only exception to this point is when the expression in the bracket in \eqref{eq:delta_dAB-Qn} vanishes. One should notice that it is not enough that the integral in that expression vanishes. Non-zero contribution comes also from the jumps of the radion fields at the branes. So, for example a radion may change distances between points on two different branes even if its profile is exactly zero in the whole section between those branes.

Typically changes in distances caused by heavier radions are smaller. There are two reasons. First: profiles of heavier radions are more rapidly oscillating functions, so there are stronger cancellations between positive and negative contributions to the integral in \eqref{eq:delta_dAB-Qn}. Second: it follows from the JC \eqref{eq:JC_Qn2} that jumps of the profiles, $[Q_{m_n^2}]_I$, decrease with increasing $m_n^2$ so also the boundary contributions to \eqref{eq:delta_dAB-Qn} are smaller for heavier radions.

\item[c)]
Expression \eqref{eq:delta_dAB-Qn} simplifies for the case of points on two end-of-the-world orbifold branes. Radion profiles are continuous at those branes, $[Q]_1=0=[Q]_2$ so only the first term in the bracket in \eqref{eq:JC_Qn2} gives (in general) non-zero contribution.

\item[d)]
The JC \eqref{eq:JC_A'} tells us that jumps of $A'$ present in eq.~\eqref{eq:delta_dAB-Qn} are proportional to tensions of the corresponding intermediate branes. Thus, the brane located contributions to $\delta d_{AB}$ are more important for branes with smaller tensions. In other words: branes with smaller tension are more easy to be deformed.

\end{itemize}

So far we have discussed distances along the extra 5th dimension. However, radions influence also distances along ordinary dimensions. Let us now consider two points $C$ and $D$ on the $j$-th brane with the coordinates $(t,x^1_C,x^2,x^3,y_j)$ and $(t,x^1_D,x^2,x^3,y_j)$. Performing calculations analogous to those for the previous case of points $A$ and $B$, but with $g_{55}$ replaced with $g_{11}$, one can find
\begin{align}
\label{eq:dCD}
d_{CD}
&\approx
e^{-A(y_j)}(x^1_D-x^1_C)
\nonumber\\
&
\quad+
e^{A(y_j)}\left[\frac{1}{A'}\right]_j^{-1}
\sum_n
\left[\frac{Q_{m_n^2}}{A'}\right]_j
\int_{x^1_C}^{x^1_D}f_{m_n^2}(t,\vec{x})\,\d x^1
+
\frac12e^{-A(y_j)}
\sum_nE_{m_n^2}(y_j)\left(\partial_{x^1}f_{m_n^2}(t,\vec{x})\right)\!\Big|_{x^1_C}^{x^1_D}.
\end{align}
In the case of the orbifold branes products of jumps present in the above formula simplify to
\begin{equation}
\left[\frac{1}{A'}\right]_{1,2}^{-1}\left[\frac{Q_{m_n^2}}{A'}\right]_{1,2}
=
Q_{m_n^2}(y_{1,2})\,.
\end{equation}
We will not discuss formula \eqref{eq:dCD} in much detail. However, its two features are worth mentioning. 
First: propagating radions, similarly as propagating gravitons, distort 4D space-time changing distances between points with fixed 4D coordinates\footnote{
For simplicity we refer to distances at fixed time but the same reasoning may be applied to relativistically invariant proper distances.}.
Very large number of gravitons may be described as gravitational waves. Analogously, very large number of radions may be interpreted as another kind of gravitational waves. The main difference is that traditional gravitational waves have tensor character (gravitons are tensor perturbations of the metric) while waves related to radions have scalar character. For this reason all running and planned experiments looking for gravitational waves can not detect such radion waves because these experiments are designed to be sensitive to tensor distortions of the space-time.
Second: the correction to $d_{CD}$ due to propagating radions is relatively more important for points on the IR brane and on branes located close to it. The unperturbed distance, term in the first line of \eqref{eq:dCD}, is proportional to the warp factor $\exp({-A})$
while the correction, terms in the second line of \eqref{eq:dCD}, is typically dominated by the contribution which scales as the inverse of the warp factor, i.e.~as $\exp({+A})$.

%%%%%%%%%%%%%%%%%%%%%%%%%%%%%%%%%%%%%%%%%%%%%%%%%%%%%%%%%%%%%%%%%%
\section{Summary}
\label{sec:summary}
%%%%%%%%%%%%%%%%%%%%%%%%%%%%%%%%%%%%%%%%%%%%%%%%%%%%%%%%%%%%%%%%%%

In 5D models with the 5th dimension compactified on the orbifold 
$S^1/\mathbb{Z}_2$ there are always two end-of-the-world orbifold branes located at both fixed points of the $\mathbb{Z}_2$ symmetry. It is possible that between those orbifold branes there are additional intermediate branes. In this kind of models there are background configurations with 
fixed positions of all branes if a Goldberger-Wise scalar field with some bulk and brane potentials is added. However, fixing positions of all branes is not necessarily enough to have a reasonable model. It is necessary that a given configuration is stable against small scalar perturbations of the background metric and perturbations of the GW field. 
There were some discussion in the literature on how many scalar perturbations of the metric must be taken into account. 
We checked that, as claimed by some authors, in multibrane models it is necessary to consider one additional perturbation as compared to models without intermediate branes. We have shown that these four scalar perturbations (three in the metric and perturbation of the GW field) reduce to only one independent combination when the equations of motion are used. This combination is a 5D scalar field so it gives rise to one infinite tower of 4D Kaluza-Klein modes. All these 4D modes we call radions. Thus, there is infinite number of radions irrespectively of the number of intermediate branes. This is in analogy to the infinite number of 4D (massless and massive) gravitons in 5D models. Gravitons are KK modes of 5D tensor perturbations of the metric while radions (sometimes called graviscalars) are KK modes of 5D scalar perturbations of the metric. The lightest graviton is massless while the smallest mass squared of radions may have either sign or may be zero.

Stability of background configurations in multibrane models is the main subject of this work. This stability depends on the sign of the lightest radion mass squared. A given configurations is stable (unstable) if the lightest radion mass squared is positive (negative). Unstable configurations are ``destroyed'' by exponential grow of tachyonic radion(s) i.e.~tachyonic KK mode(s) of scalar perturbations. Limiting cases with massless radions (and no tachyonic ones) are more complicated and require calculations of higher order in small 
perturbations\footnote{
Similarly as in mechanics where the second derivative of a potential may vanish at local minimum or maximum or may even correspond to a flat potential. One example is the original RS 2-brane model without GW scalar in which (the only) radion is massless.  Positions of the branes in this model are not fixed in analogy to neutral mechanical equilibrium.}.

The spectrum of radions consists of all eigenvalues of the system describing their dynamics. This system consists of appropriate bulk equations of motion, boundary conditions at orbifold branes and junction conditions at intermediate branes. We showed that the bulk EoM may be written in the form of Sturm-Liouville equation analogous to that for the two brane models though the radion field depends also on additional scalar perturbation of the metric $E$. We found the most general (without very often used stiff brane potential approximation) form of junction conditions at intermediate branes \eqref{eq:JC_Qn1} and \eqref{eq:JC_Qn2}.

Analyzing the obtained EoM and JCs we found the set of conditions to have a given sign of the smallest radion mass squared. When this sign is positive all radions have positive masses (squared) and the corresponding background is stable. We repeat here the necessary and sufficient conditions for this to happen.
\begin{enumerate}
\item
All brane parameters $B_j$ are positive:
\begin{equation*}
B_j=e^{-A(y_j)}U^{\prime\prime}_j
\big(\phi(y_j)\big)-\left[A'\right]_j-\left[\frac{\phi^{\prime\prime}}{\phi'}\right]_j>0\,.
\end{equation*}
\item
Derivative of the GW scalar background is nowhere vanishing:
\begin{equation*}
\begin{cases}
\hspace{42pt}
\phi'(y)\ne0 \qquad \text{in the bulk} 
\\
\lim_{y\to y_j^\pm}\phi'(y)\ne0 \qquad \text{at all branes} 
\end{cases}\,.
\end{equation*}
\item
All intermediate branes $I_i$ have non-negative tension resulting in non-negative jumps of $A'$:
\begin{equation*}
\left[A'\right]_{I_i} \ge0 \,.
\end{equation*}
\end{enumerate}
A given multibrane configuration is stable when the above conditions are fulfilled. 
The necessary and sufficient condition leading to at least one massless radion was obtained in Section \ref{sec:m=0} 
%and are given in eq.~\eqref{eq:condition_for_m=0} 
and reads
\begin{equation*}
\prod_j B_j\phi'(y_j^-)\phi'(y_j^+)=0\,.
\end{equation*}
The conditions complementary to those listed above (i.e.:~at least one $B_j<0$ or at least one $\left[A'\right]_{I_i} <0 $ or $\phi'=0$ anywhere in the bulk) are necessary and sufficient for the existence of at least one  tachyonic radion, so leading to instability of a given multibrane system.

Formulation of all the above mentioned conditions related to stability of multibrane configurations is the main result of the present work. 

All the presented above results apply also to any model with only two branes. In such cases they simplify considerable and agree with the known results for 2-brane models. The main simplifications of 2-brane models are: smaller number of scalar perturbations to be considered; one simple BC at each brane instead of two more complicated JCs; no need for the condition on the intermediate branes tensions (point 3 above); no discontinuities of the radion fields.

In addition some properties of radions were discussed in Section \ref{sec:properties}. If a radion is massless in a given model for some values of the parameters then it may be unusually light when some of those parameters are slightly changed. It was shown that in a model with $N$ branes there may be up to $2^{N-1}$ massless and exceptionally light radions. Possibilities of radions with profiles exactly vanishing in some of the bulk sections (assumed to exist by some authors) were discussed. It was shown for example that this is possible for a massless radion only if another massless radion exists in the model. Finally the relations between radions and distances were discussed. Expressions relating radions excitations to distances between branes but also along branes were found and discussed. Propagating radions, similarly as propagating gravitons, deform the space-time and may be associated with some kind of gravitational waves. Such waves have scalar character so are different from the standard gravitational waves causing tensor deformation of space-time. Presented in this work relations between radions and distances should be contrasted to often used in the literature description in which distance(s) between branes are given by vacuum expectation values of radion(s). Radions, being the 4D modes of scalar perturbations, by definition have vanishing VEVs. The distances between branes are fixed by solving appropriate systems of bulk EoMs and JCs.

%%%%%%%%%%%%%%%%%%%%%%%%%%%%%%%%%%%%%%%%%%%%%%%%%%%%%%%%%%%%%%%%%%
\section*{Acknowledgments}
Work partially supported by National Science Centre,
Poland, grant DEC-2018/31/B/ST2/02283. 
I would like to thank Stefan Pokorski for many discussions.
%%%%%%%%%%%%%%%%%%%%%%%%%%%%%%%%%%%%%%%%%%%%%%%%%%%%%%%%%%%%%%%%%%

%%%%%%%%%%% BIBLIOGRAPHY %%%%%%%%%%%%%%%%%
\bibliographystyle{BiblioStyle}
\bibliography{Multibrane_stability}

%%%%%%%%%%%%%%%%%%%%%%%%%%%%%%%%%%%%%%%%%%%%%%%%%%%%%%%%%%%
%%%%%%%%%%%%%%%%%%%%%%%%%%%%%%%%%%%%%%%%%%%%%%%%%%%%%%%%%%%
\end{document}